\documentclass[amsmath,amssymb,twocolumn,showpacs,aps,prb,10pt]{revtex4-2} 

\usepackage{graphicx}
\usepackage{ifthen}
\usepackage{color}
\usepackage{bm}
\usepackage{braket}
\usepackage{multirow}
\usepackage{hyperref}
\usepackage{soul}
\hypersetup{
 pdfnewwindow=true, colorlinks=true,
 linkcolor=blue, anchorcolor=blue,
 citecolor=blue, filecolor=blue,
 menucolor=blue, urlcolor=blue}
\usepackage{booktabs} 
\usepackage{longtable} 

\begin{document}

\title{Inverse Faraday effect in 3$d$, 4$d$, and 5$d$ transition metals} 

\author{Shashi B. Mishra}
\email{mshashi125@gmail.com}
\affiliation{Mechanical Engineering, University of California, Riverside, California 92521, USA.}
\thanks{Current address: Department of Physics, and Astronomy, Binghamton University-SUNY, Binghamton, New York 13902, USA}
\date{\today}

\begin{abstract}

Using first-principles calculations, we systematically investigate the spin contributions to the inverse Faraday effect (IFE) in transition metals. The IFE depends on the $d$-electron filling and asymmetry between excited electron and hole spin moments. Our results reveal that even elements with smaller electron magnetic moments, like Os, can exhibit higher IFE due to greater electron-hole asymmetry. Pt shows the highest IFE in the 1$–$2~eV frequency range, while Os dominates in the 2$–$4~eV range. In addition, we demonstrate that the IFE of neighboring elements with similar crystal structures (e.g., Ir, Pt, and Au) can be tuned by adjusting their Fermi levels, indicating the importance of $d$ electron filling on IFE. Finally, we find that the trend in electron (or hole) contributions to the IFE closely follows that of the spin Hall conductivity (SHC), however, the total IFE involves more complex interactions. 
\end{abstract}

\maketitle

\section{\label{sec:intro}Introduction}

The inverse Faraday effect (IFE) is a nonlinear optical phenomenon wherein circularly polarized light (CPL) induces a magnetic moment in a material without requiring an external magnetic field~\cite{van1965,Pershan1966}. This opto-magnetic coupling enables all-optical helicity-dependent switching (AO-HDS)~\cite{Mangin_2012,Lambert2014,Mangin2014}. Experimentally, IFE has been demonstrated to reverse the magnetization in ferrimagnets with a strong laser pulse~\cite{Kimel2005,Stanciu2007}, and to generate light-induced torques on magnetic metals~\cite{Choi2017}. Given these capabilities, IFE holds a key mechanism for applications in ultrafast magnetism~\cite{RevModPhys2010_Rasing} and magneto-optics~\cite{Cheng2020,Ortiz2023}.

The IFE was first described phenomenologically in the 1960s~\cite{Pitaevskii1961,Pershan1963}. Later, it was revisited, and a model for IFE in an isotropic, collisionless electron plasma was derived~\cite{HERTEL2006L1}. To better understand the behavior in real materials, several quantum mechanical theories of IFE have been proposed, particularly for magnetic metals~\cite{Berritta2016,Freimuth2016,Scheid2019,Scheid_2021}. However, for non-magnetic metals with inversion symmetry, where the electronic bands are doubly degenerate, special attention is needed. In our previous work~\cite{Shashi2023}, we developed a gauge-invariant IFE theory that accounts for the induced spin contributions in such systems. 

In this study, we extend this framework to investigate the spin contributions to the IFE across 3$d$, 4$d$, and 5$d$ transition metals. Since most of these elements are non-magnetic, except Fe, Co, and Ni, we consider hypothetical non-magnetic configurations for these three elements to enable a consistent comparison across the series. This approach facilitates a systematic exploration of how the IFE evolves with the number of valence electrons and helps reveal the underlying material properties that govern the effect. The complex case of magnetic systems, where intraband contributions from photoexcited electron-hole pairs become significant, will be addressed in future work. 

Recent studies have highlighted trends in the spin Hall effect (SHE) across transition metals~\cite{Tanaka2008, Du2014, Jo2018, Salemi2022,Go2024}.
As one moves across a given row of the periodic table, the number of valence electrons increases. Conversely, moving down the table from 3$d$ to 5$d$ elements increases both spin-orbit interaction (SOI) strength and $d$-band width, even for elements with the same number of outer valence electrons. The SHE is known to scale with both the number of valence electrons and SOI strength. Despite these insights, a systematic study of how the IFE varies with valence electron count across transition metals remains lacking.

To fill this gap, we analyzed the IFE in 30 transition metals using the theoretical expression given in Eq.~\eqref{eq:ife_all_three}, which accounts for contributions from both excited electron and hole moments. We find that the trend in doubly-resonant electron contributions to IFE closely aligns with that of the SHC, where a transverse electric field generates a spin current~\cite{Tanaka2008, Sinova2015}. However, the total IFE depends on intricate factors, particularly the subtle asymmetries between the magnetic moments of excited electrons and holes. These asymmetries are influenced by SOI, differences in effective masses between electrons and holes~\cite{Fuseya2015}, and the intrinsic band structures of the materials~\cite{Karplus1954,Tanaka2008}.  

Among the systems studied, Pt exhibits the highest spin IFE in the 1$–$2~eV frequency range, while Os shows the strongest IFE response in the 2$–$4~eV range, but with an opposite sign to that of Pt. To elucidate the origin of Pt’s large IFE, we demonstrate that the IFE of its neighboring elements, Ir and Au, can be made comparable by tuning their Fermi levels. Such tuning—achievable via metal doping or alloying—offers a practical route to engineer the IFE response in transition metals.

The rest of the paper is organized as follows: Sec.~\ref{sec:theory} provides a brief overview of the IFE. Sec.~\ref{sec:methods} outlines the computational methods used. The main results are presented in Sec.~\ref{sec:results}, followed by detailed discussion in Sec.~\ref{sec:discussion} and summary in Sec.~\ref{sec:summary}. Additional details are provided in the Appendix.

\begin{figure*}[!hbt]
    \centering
    \includegraphics[scale=0.85]{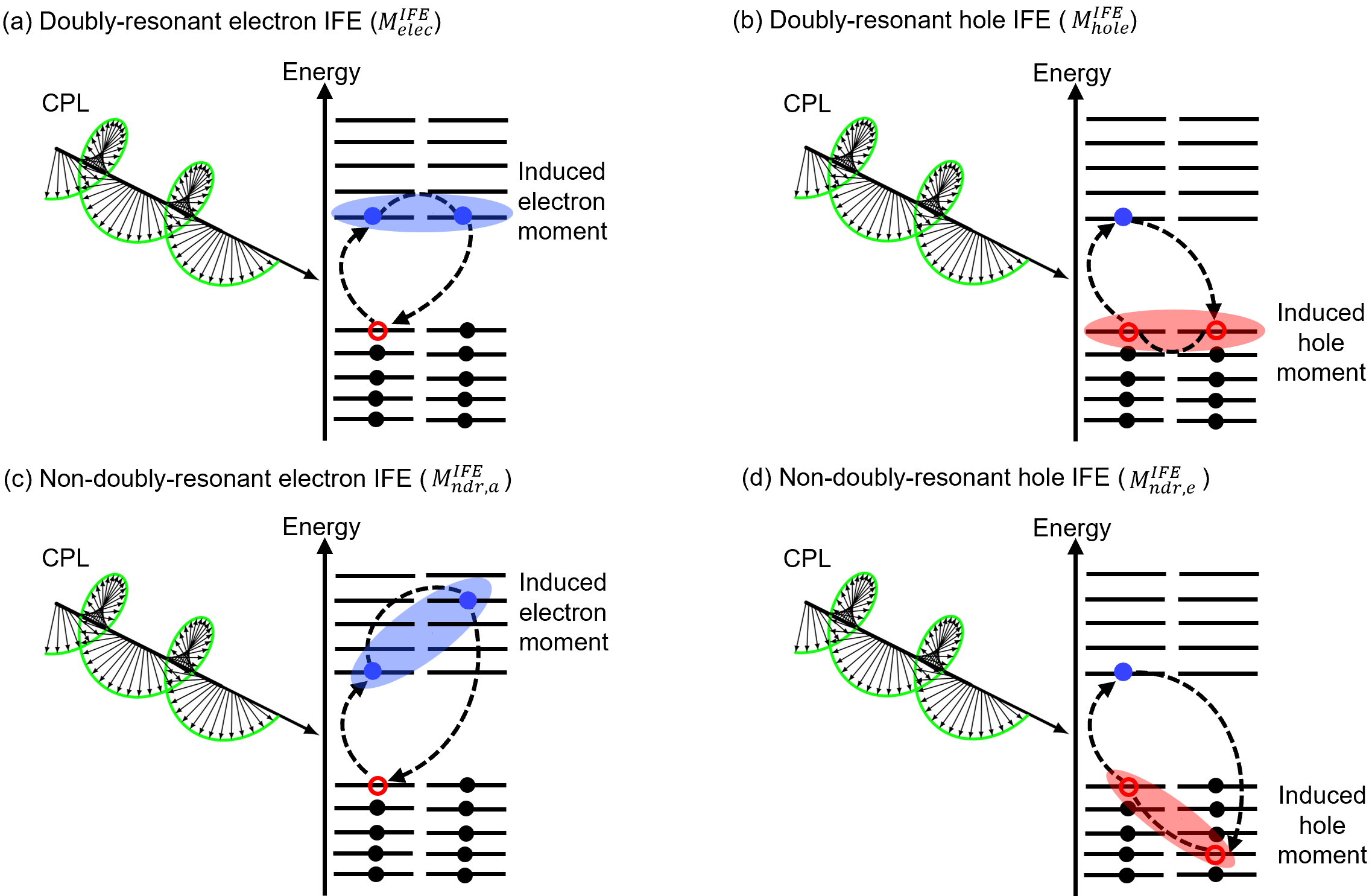}
    \caption{\label{fig:schematic_magn} Schematic illustration of the dominant contributions to the IFE in metals with doubly degenerate band structures. Band doublets are shown as pairs of solid lines at each energy level. Panels (a) and (b) depict the doubly-resonant contributions to the IFE from excited electron ($M^{\rm IFE}_{\rm elec}$) and hole ($M^{\rm IFE}_{\rm hole}$) moments, respectively. Panels (c) and (d) illustrate the dominant non-doubly-resonant contributions from electrons ($M^{\rm IFE}_{\rm ndr,a}$) and holes ($M^{\rm IFE}_{\rm ndr,e}$), respectively.}
\end{figure*}

\section{\label{sec:theory}Theory}

The IFE is a second-order response of a material to an external electromagnetic field. Its theory has been derived using either the density matrix approach~\cite{Battiato2014} or a perturbative formalism~\cite{Popova2011,Shashi2023}. Both methodologies calculate the IFE by summing over occupied (valence) and empty (conduction) states, a technique known as the sum-over-states method. A key consideration in this approach is the treatment of band degeneracy.  For instance, in nonmagnetic metals with inversion symmetry, the bands are doubly degenerate~\cite{Elliott1954}, corresponding to spin-up and spin-down states in the absence of SOI. With SOI, these states are still degenerate but cannot be labeled as pure spin states. The sum-over-states method poses challenges when dealing with such doubly degenerate bands~\cite{Pientka2012}, particularly in the second-order perturbation calculations of the IFE~\cite{Shashi2023}. To systematically address this, we define each electronic state using a band index $n$ and a spinor-index $N$, where \(N=1,2\) distinguishes the two components of the degenerate doublet, corresponding to the same eigenenergy $E_{n\bm{k}}=E_{nN\bm{k}}$, with $\bm k$ as the crystal momentum. The Bloch states for these doublets are denoted as $\ket{\phi_{nN\bm{k}}}$. 

The IFE for such doubly degenerate bands can be decomposed into three distinct terms~\cite{Shashi2023}:  
\begin{equation}\label{eq:ife_all_three}
    M^{\rm IFE} =  M^{\rm IFE}_{\rm elec}- M^{\rm IFE}_{\rm hole} + M^{\rm IFE}_{\rm ndr}.
\end{equation}
Here, $M^{\rm IFE}_{\rm elec}$ and $M^{\rm IFE}_{\rm hole}$ represent doubly resonant contributions from excited electron and hole spin moments, respectively. These terms are expressed through the integrals over the first Brillouin zone (BZ) as follows:
\begin{widetext}
\begin{align}
M^{\rm IFE}_{\rm elec} =&  
\int_{\rm BZ} 
\frac{d^3k}{(2\pi)^3} 
\sum_n^{\rm occ} 
\sum_{N=1}^2
\sum_m^{\rm emp}
\sum_{M=1}^2
\sum_{M'=1}^2 
\frac{
\braket{\phi_{nN\bm{k}} | V | \phi_{mM\bm{k}}} 
\braket{\phi_{mM\bm{k}} | M^{\rm spin} | \phi_{mM'\bm{k}}}
\braket{\phi_{mM'\bm{k}} | V^{\dagger} | \phi_{nN\bm{k}}} }
{(E_{m\bm{k}} - E_{n\bm{k}} - \hbar \omega)^2 + \eta^2},
\label{eq:ife_elec}
\\
M^{\rm IFE}_{\rm hole} = & 
\int_{\rm BZ} 
\frac{d^3k}{(2\pi)^3} 
\sum_n^{\rm occ} 
\sum_{N=1}^2
\sum_{N'=1}^2
\sum_m^{\rm emp}
\sum_{M=1}^2
\frac{
\braket{\phi_{nN\bm{k}} | V | \phi_{mM\bm{k}}} 
\braket{\phi_{mM\bm{k}} |  V^{\dagger} | \phi_{nN'\bm{k}}}
\braket{\phi_{nN'\bm{k}} | M^{\rm spin} | \phi_{nN\bm{k}}} }
{(E_{m\bm{k}} - E_{n\bm{k}} - \hbar \omega)^2 + \eta^2}.
\label{eq:ife_hole}
\end{align}
\end{widetext}

The electric field of the incident light is assumed to be in the $x$-$y$ plane, inducing the magnetic moment along the $z$-direction. The spin matrix elements are given by,
$\braket{\phi_{mM\bm{k}} | M^{\rm spin} | \phi_{mM'\bm{k}}} 
=2 \frac{e}{2 m_{\rm e}} 
\braket{\phi_{mM\bm{k}} | S_z | \phi_{mM'\bm{k}}}$, where \(S_z\) is the spin operator along the \(z\)-axis. The interband transition matrix ($n\neq m$) elements are expressed as   
$\braket{\phi_{nN\bm{k}} | V | \phi_{mM\bm{k}}} 
=\frac{e}{2}\sqrt{\frac{I}{\epsilon_0 c}} 
\frac{E_{n\bm{k}} - E_{m\bm{k}}}{\hbar \omega}
\left( A_{nNmM\bm{k}}^x + i A_{nNmM\bm{k}}^y \right)$, where $A_{nNmM\bm{k}}^{\alpha}$ is the Berry connection matrix, $e$ is the charge of the electron, $I$ is the intensity of the incoming light, $c$ is the speed of light, $\epsilon_0$ is the permittivity of the free space, and $\eta$ is the lifetime of the excited electron.  Further details on the derivation can be found in Ref.~[\onlinecite{Shashi2023}].

In Eq.~\eqref{eq:ife_elec}, the sums over $N$, $M$, and $M'$ run over the spinor components of the doubly degenerate bands at each \(\bm{k}\)-point. A resonance occurs when the energy difference between an occupied state \(n\) and an unoccupied state \(m\) satisfies \(E_{m\bm{k}} - E_{n\bm{k}} \approx \hbar \omega\), corresponding to the absorption of a photon. Since the IFE is a second-order nonlinear optical process, it involves a second optical transition. If this second transition occurs within the spinor doublets, where the states are nearly degenerate in energy, the excitation and de-excitation levels are energetically matched, leading to a double-resonance condition. Under these circumstances, the dominant contributions from such doubly resonant processes are captured by \(M^{\rm IFE}_{\rm elec}\) and \(M^{\rm IFE}_{\rm hole}\), as schematically illustrated in Figs.~\ref{fig:schematic_magn}(a) and~\ref{fig:schematic_magn}(b), respectively.

In contrast, if the second transition involves states that are not energetically close, such transitions between different bands or non-spinor-degenerate states, the contribution is classified as non-doubly-resonant, and is represented by \(M^{\rm IFE}_{\rm ndr}\). In these cases, only one denominator becomes resonant while the other remains finite for a given photon energy $\hbar \omega$, resulting in a weaker contribution compared to the double-resonant terms unless the numerator (product of matrix elements) varies significantly. The full expression for \(M^{\rm IFE}_{\rm ndr}\), derived in our previous work~\cite{Shashi2023}, consists of eight distinct terms, which are summarized in Appendix~\ref{app:nonresonant}. Among these, the two dominant terms, \(M^{\rm IFE}_{\rm ndr,a}\) and \(M^{\rm IFE}_{\rm ndr,e}\), are schematically illustrated in Figs.~\ref{fig:schematic_magn}(c) and~\ref{fig:schematic_magn}(d), respectively.

\section{\label{sec:methods}Computational Methods}

We performed density functional theory calculations using Quantum {\small ESPRESSO}~\cite{Giannozzi2017} within the generalized gradient approximation~\cite{Perdew_1996}. We used fully relativistic ONCV pseudopotentials~\cite{Hamann2013} from the pseudo-dojo library~\cite{VANSETTEN201839}. The kinetic energy cutoff for the plane wave basis expansion was set to 120~Ry. 

For all transition metals studied, we used experimental lattice parameters. Specifically, we considered body-centered cubic (bcc) structures for V, Cr, Mn, Fe, Nb, Mo, Ta, and W, and face-centered cubic (fcc) structures for Ni, Cu, Rh, Pd, Ag, Ir, Pt, and Au. The bcc and fcc structures contain one atom in the primitive lattice. For hexagonal-close-packed (hcp) metals, including Sc, Ti, Co, Y, Zr, Tc, Ru, Lu, Hf, Re, Os, Zn, and Cd, we used two atoms in the primitive cell. For Hg, we considered the solid rhombohedral structure with a two-atom primitive cell. 

To investigate Fe, Co, and Ni in their nonmagnetic states, we performed non-spin-polarized DFT calculations, treating these elements as paramagnetic metals. Their respective room-temperature crystal structures- bcc for Fe, hcp for Co, and fcc for Ni were used. In these calculations, we set the initial magnetic moments to zero so that the magnetization was suppressed during the self-consistent field (SCF) iterations. Spin-orbit coupling was included to study relativistic effects while preserving time-reversal symmetry. This approach allows for a direct comparison with other transition metals in a hypothetical nonmagnetic state.

We performed the SCF calculations using a \(24 \times 24 \times 24\) \(\mathbf{k}\)-mesh for fcc and bcc structures, and a \(16 \times 16 \times 16\) mesh for hcp metals. Subsequent non-self-consistent field (NSCF) calculations were carried out on a coarser \(8 \times 8 \times 8\) \(\mathbf{k}\)-grid to construct maximally localized Wannier functions (MLWFs) using Wannier90~\cite{Pizzi2020, Marzari2012}. To test convergence, we also examined NSCF calculations with denser \(\mathbf{k}\)-meshes of \(12 \times 12 \times 12\) and \(16 \times 16 \times 16\) for Pt as a representative system (see Fig.~\ref{fig:grid-conv} in Appendix~\ref{app:grid-conv}). The results indicate that the spin IFE magnitude remains effectively unchanged with these finer NSCF grids. Additionally, a lower SCF \(\mathbf{k}\)-mesh of \(12 \times 12 \times 12\) was found to be accurate for achieving convergence of charge densities in the case of Pt, while it slightly differs for Au. Therefore, for consistency across all fcc and bcc structures, we employed the denser \(24 \times 24 \times 24\) SCF \(\mathbf{k}\)-mesh in our calculations. 

For fcc and bcc metals, 18 atom-centered Wannier orbitals-comprising \(sp^3d^2\), \(d_{xy}\), \(d_{yz}\), and \(d_{zx}\) were constructed, while for hcp metals with two atoms in the primitive cell, 36 Wannier functions were projected. The spread of the \(sp^3d^2\)-projected orbitals was approximately \(1.16\) \AA$^2$, and that of the \(d_{xy}\), \(d_{xz}\), and \(d_{yz}\) orbitals was about $\sim$\(0.81\) \AA$^2$, with an average total spread of \(1.04\)$\sim$\AA$^2$. Wannier interpolation~\cite{Wang2006} was performed using Wannier~Berri program~\cite{Stepan2021}.

\begin{figure*}[!t]
    \centering
    \includegraphics{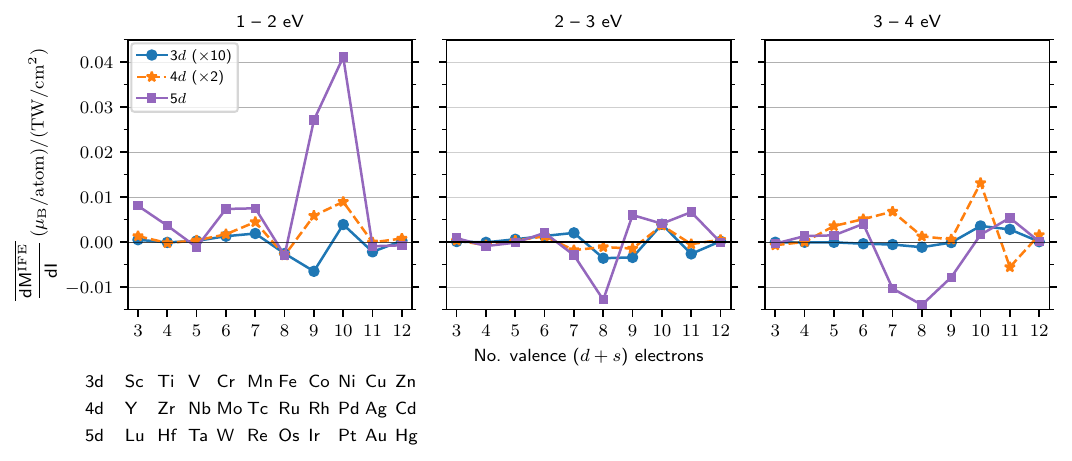}
    \caption{\label{fig:mean-ife-total} Average spin IFE of 3$d$, 4$d$, and 5$d$ metals as a function of the number of valence ($d$+$s$) electrons. The IFE values are averaged over three frequency ranges: 1$-$2 eV (left), 2$-$3 eV (middle), and 3$-$4 eV (right). The full frequency dependence of IFE is given in Figs.~\ref{fig:3d-metals}$-$\ref{fig:5d-metals} of the Appendix~\ref{app:IFE-with-freq}.  For clarity, the IFE for 3$d$ and 4$d$ metals are scaled by factors of 10 and 2, respectively.}
\end{figure*}

We found that a \textbf{k}-point interpolation grid of $100\times100\times100$ was sufficient to obtain convergence for fcc and bcc metals, while for hcp structures, we used a $70\times70\times70$ \textbf{k}-mesh. To avoid singular points in the Brillouin zone arising from the high symmetry points, we applied a small random shift to the uniform interpolation grid along all three cartesian directions. A constant inverse lifetime ($\eta$) of 0.1 eV was used in all calculations to account for broadening effects. The spin hall conductivity (SHC) was calculated for all metals assuming electric fields in the $x-y$ plane so that the spin polarization was along the $z$ direction. These calculations were performed at $\omega$ = 0 with $\eta$ = 0.1~eV using Wannier~Berri package~\cite{Qiao2018}.

\section{\label{sec:results}Results}

\subsection{\label{IFE-freq} Frequency dependence of IFE in \texorpdfstring{3$d$, 4$d$, and 5$d$}{3d, 4d and 5d} metals}

The IFE depends on the frequency of light, and it can be decomposed into doubly-resonant (\( M^{\rm IFE}_{\rm elec} - M^{\rm IFE}_{\rm hole} \)) and non-doubly-resonant (\( M^{\rm IFE}_{\rm ndr} \)) terms, as discussed in Sec.~\ref{sec:theory}. The frequency dependence of these components for 3$d$, 4$d$, and 5$d$ metals is shown in Figs.~\ref{fig:3d-metals}$-$\ref{fig:5d-metals} of Appendix~\ref{app:IFE-with-freq}. For most transition metals—particularly the 5$d$ elements, where SOI is stronger—the doubly-resonant terms dominate the overall spin IFE. In contrast, for metals with fully filled valence shells, such as Zn, Cd, and Hg, the doubly- and non-doubly-resonant contributions are similar in magnitude but opposite in sign, resulting in a near cancellation of the total spin IFE. This behavior is attributed to the reduced asymmetry between electron and hole spin moments in systems with fully occupied outer $d$ and $s$ states, which reduces the magnitude of the doubly resonant term. Similar observations for filled orbitals have been reported in the literature for spin-Hall conductivity (SHC) calculations~\cite{Salemi2022,Go2024}. Furthermore, we find that group 11 elements, such as Cu, Ag, and Au, exhibit spin IFE peaks in their respective interband optical resonance regions~\cite{Theye_1970,Christensen_1971,Uba2017}.

\subsection{\label{IFE-results} Average IFE as a function of valence electrons}

To analyze trends in the spin IFE across the transition metal series, we computed the IFE response at discrete photon energies from 1 to 4~eV, using a frequency step of 0.02~eV. We then extracted average IFE values within three-photon energy intervals: 1$-$2~eV, 2$-$3~eV, and 3$-$4~eV. These intervals span the near-infrared to visible and extend into the near-ultraviolet region, which is relevant for most practical applications. The resulting averages are shown in Fig.~\ref{fig:mean-ife-total}. Across the 3$d$, 4$d$, and 5$d$ transition metals, the IFE exhibits broadly similar dependence on the total number of valence ($d$+$s$) electrons. However, the overall magnitude increases systematically with the atomic number due to enhanced SOI~\cite{Sarma1981}, with 5$d$ metals showing the strongest responses. For example, in the 1$-$2~eV range, the spin IFE of Pt is higher than that of Pd, following an approximate \(Z^4\) scaling of SOI. IFE values are also enhanced for elements with 8, 9, and 10 valence electrons, indicating the importance of $d$-orbital occupancy near the Fermi level.

While these trends are clear, establishing a universal scaling of IFE with atomic number is challenging due to the complex structure of the underlying expression (Eq.~\eqref{eq:ife_all_three}). As the photon energy increases from 1$-$2~eV to 2$-$4~eV, the IFE is influenced by competing factors: a \(1/\omega^2\) prefactor (see Eq.~17 of Ref.~\onlinecite{Shashi2023}) that suppresses high-energy response, and the emergence of material-specific interband resonances. For instance, Os and Au exhibit strong $d \rightarrow s$ interband transitions above 2~eV (see Fig.~\ref{fig:5d-metals} in Appendix~\ref{app:IFE-with-freq}), leading to a significant enhancement of the IFE in the higher energy range.

\begin{figure*} [!hbt]
    \centering
    \includegraphics[scale=0.9]{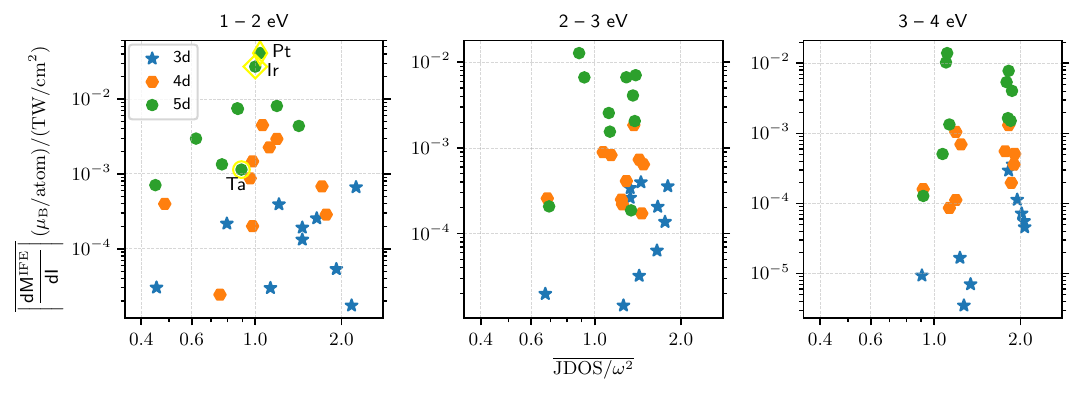}
    \caption{\label{fig:mean-jdos-tot} Relationship between the mean absolute value of IFE and JDOS$/\omega^2$ for 3$d$, 4$d$ and 5$d$ metals in three frequency ranges 1$-$2 eV (left), 2$-$3 eV (middle) and 3$-$4 eV (right). The vertical axis for mean IFE values is shown on a logarithmic scale.} 
\end{figure*}

\begin{figure*}[!ht]
    \centering
    \includegraphics{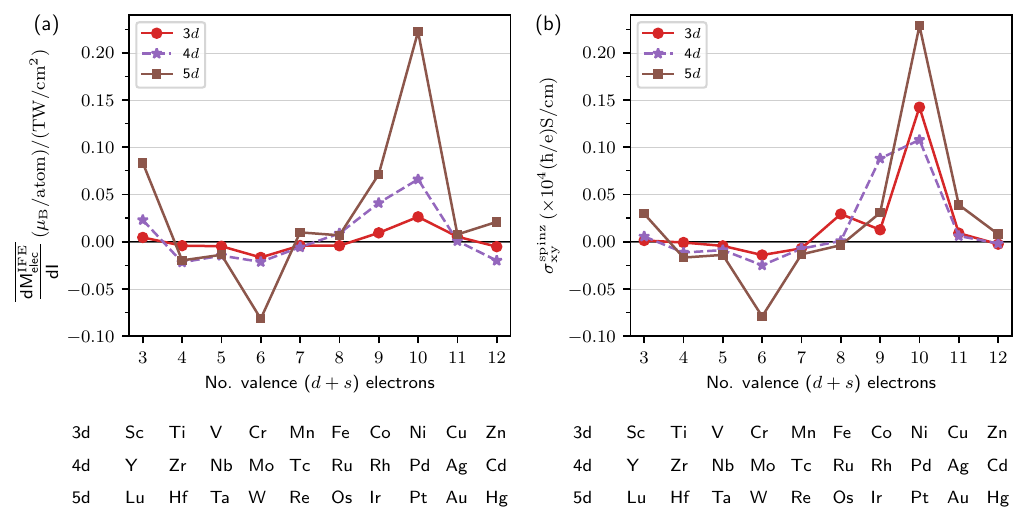}
    \caption{\label{fig:ele-1ev} (a) Calculated mean value of $M^{\rm IFE}_{\rm elec}$ for 3$d$, 4$d$ and 5$d$ transition metals in the 1$-$2~eV frequency range, plotted as a function of valence electron filling. (b) The real part of spin hall conductivity (SHC) $\sigma_{\rm xy}^{\rm spin z}$ as a function of valence electrons. The SHC is computed at zero frequency ($\omega$=0) and at each metal's Fermi energy, using a broadening parameter $\eta$=0.1~eV.}
\end{figure*}

Based on the observed IFE trends, we categorize the 30 transition metals into four categories. The first includes early transition metals, which have one to three $d$ electrons. Their IFE values decrease steadily from group 3 to 5 in the 1$–$2~eV range. The second category contains groups 6 and 7, with nearly half-filled $d$ orbitals. These metals exhibit stronger IFE responses than the earlier ones, particularly in the case of Cr and Mo, where doubly-resonant electron ($M^{\rm IFE}_{\rm elec}$) and hole terms ($M^{\rm IFE}_{\rm hole}$) are large (Fig.~\ref{fig:ele-1ev}).

Next are elements with six to eight $d$ electrons (groups 8$-$10). 
Those with six (e.g., Fe, Ru, and Os) tend to show negative IFE values, with Os having the largest IFE magnitude among all metals in the 2$–$4~eV range. In contrast, elements like Ni, Pd, and Pt—having nearly filled $d$ shells—exhibit strong positive IFE in the 1$–$2~eV regime, with Pt showing the highest overall response due to its strong SOI and electronic configuration.

The final category includes groups 11 and 12, characterized by filled $d$ bands and conduction in $s$ states. In Cu, Ag and Au, the IFE increases near the $d \rightarrow s$ interband transition region, with Au showing a strong enhancement above 2~eV, and Ag has a peak around 3~eV, both driven by interband optical resonances~\cite{Theye_1970,Christensen_1971,Ortiz2022,Uba2017} (see Appendix~\ref{app:IFE-with-freq}). On the other hand, Zn, Cd, and Hg exhibit minimal IFE, due to cancellation between doubly resonant and non-doubly resonant terms caused by fully filled $d$ and $s$ states. 

To further interpret these trends, we analyze key factors such as the number of optical transitions, the magnitude of the doubly resonant terms, and the SOI-driven asymmetry in excited electron-hole spin moments, as discussed in the following sections.

\section{\label{sec:discussion}Discussion}
\subsection{Analysis of average IFE with \texorpdfstring{JDOS/$\omega^2$}{JDOS over omega squared}}

The IFE describes the magnetic moments induced by the interaction of excited electrons or holes with CP light. Since optical transitions from occupied to unoccupied states drive the response, one might expect the IFE to scale with the joint density of states (JDOS)(Fig.~\ref{fig:schematic_magn}). In our previous work~\cite{Shashi2023}, we showed that for a simple metal like fcc Au, the frequency dependence of the doubly-resonant electron and hole contributions ($M^{\rm IFE}_{\rm elec}$ and $M^{\rm IFE}_{\rm hole}$) resembles JDOS$/\omega^2$ behavior. 

To examine this relationship more broadly, we plot the mean absolute value of the IFE against the mean JDOS$/\omega^2$ across different frequency regimes in Fig.~\ref{fig:mean-jdos-tot}. The results reveal no clear correlation between these quantities across elements within each energy range. For instance, in the 1$-$2~eV regime, three representative 5$d$ metals—Ta, Ir, and Pt—exhibit similar JDOS values, yet the IFE of Ir and Pt is nearly an order of magnitude larger than that of Ta. Similar trends are observed across the 3$d$, 4$d$, and 5$d$ series, indicating that a higher JDOS alone does not necessarily lead to a stronger IFE.

To better understand these discrepancies, we analyze the individual doubly-resonant components—\( M^{\rm IFE}_{\rm elec} \), \( M^{\rm IFE}_{\rm hole} \), and their difference, which reflects the spin asymmetry—in the following sections. While we focus on the 1$-$2~eV range for clarity, the same approach can be extended to the 2$-$3~eV and 3$-$4~eV intervals.

\subsection{\label{sec:doubly-resonant} Correlation between \texorpdfstring{$M^{\rm IFE}_{\rm elec}$}{ele-IFE} and SHC}

Figure~\ref{fig:ele-1ev}(a) shows the mean value of the doubly-resonant electron contribution to the IFE, $M^{\rm IFE}_{\rm elec}$, averaged over the 1$-$2~eV frequency range, as a function of valence electron count. $M^{\rm IFE}_{\rm elec}$ is a frequency-dependent optical response, governed by transient electron-hole excitations under illumination. To capture typical optical behavior, we calculate $M^{\rm IFE}_{\rm elec}$ across a range of photon energies and report the average value over 1–2~eV. Across the 3$d$, 4$d$, and 5$d$ series, the overall trend is consistent, with 5$d$ metals exhibiting larger magnitudes due to stronger SOI. The sign of $M^{\rm IFE}_{\rm elec}$ varies systematically with the number of $d$ electrons ($n_{\rm d}$): except for group 3 elements, metals with \( n_d \leq 5 \) generally show negative values, while those with \( n_d > 5 \) tend to exhibit positive values. Notably, group 6 elements (Cr, Mo, and W), which have half-filled $d$ shells, show high negative values of $M^{\rm IFE}_{\rm elec}$, while group 10 metals (Ni, Pd, and Pt) exhibit large positive values, with Pt showing the highest $M^{\rm IFE}_{\rm elec}$ across all elements considered. This dependence on $d$-electron filling closely aligns the well-known behavior of spin Hall conductivity (SHC) as a function of valence electrons~\cite{Tanaka2008,Du2014,Jo2018,Salemi2022,Go2024}, motivating a direct comparison between our calculated average $M^{\rm IFE}_{\rm elec}$ and SHC values.

The SHC describes the generation of a transverse spin current in response to an applied electric field~\cite{Sinova2015}. In our IFE study, we consider an electric field applied in the \(xy\)-plane, which induces spin polarization along the \(z\)-direction. The SHC behaves similarly: an in-plane electric field gives rise to a spin current polarized along \(z\), described by the tensor component \( \sigma_{\rm xy}^{\rm spin\,z} \). According to the Kubo formalism, the SHC is given by~\cite{Yao2005,Guo2008,Qiao2018}:
\begin{align}\label{eq:SHC}
    \sigma_{\rm xy}^{\rm spin z} = 
    & \hbar \int_{\rm BZ} 
    \frac{d^3k}{(2\pi)^3}
    \sum_n f_{n\bm{k}}\nonumber \\
    \times  \sum_{ m \neq n} 
    & \frac{2 \, {\rm Im} \left[ \braket{\phi_{n\bm{k}} | \hat{j}_{\rm x}^{\rm spin z} | \phi_{m\bm{k}}} 
    \braket{\phi_{m\bm{k}} | -{\rm e}\hat{v_{\rm y}} | \phi_{n\bm{k}}}
    \right] } {(E_{m\bm{k}} - E_{n\bm{k}})^2 - (\hbar \omega + i \eta)^2}.
\end{align}
where $f_{n\bm{k}}$ is the Fermi-Dirac distribution function, $\hat{j}_{\rm x}^{\rm spin z} = \frac{1}{2} \{\hat{s}_z, \hat{v}_x\}$ is the spin current operator. 

We computed the SHC for 3$d$, 4$d$, and 5$d$ transition metals at their respective Fermi energies and at zero frequency ($\omega=0$), using a finite broadening parameter $\eta$ = 0.1~eV for consistency with our IFE calculations (see Fig.~\ref{fig:ele-1ev}(b)). A comparison between Fig.~\ref{fig:ele-1ev}(a) and Fig.~\ref{fig:ele-1ev}(b) shows that the SHC exhibits a trend similar to that of $M^{\rm IFE}_{\rm elec}$ as a function of valence electron filling. In particular, group 10 elements (Ni, Pd, and Pt) exhibit strong positive SHC values, while group 6 elements (Cr, Mo, and W) show pronounced negative peaks, consistent with previous reports~\cite{Jo2018,Salemi2022,Go2024}.

Although \(M^{\rm IFE}_{\rm elec}\) and the SHC exhibit similar trends, the underlying physical mechanisms differ fundamentally. The IFE is a frequency-dependent optical response driven by transient electron-hole excitations, while the SHC arises from steady-state Berry curvature and scattering effects~\cite{Sinova2015}. The similarity holds specifically for the resonantly enhanced electron and hole contributions (\(M^{\rm IFE}_{\rm elec}\) and \(M^{\rm IFE}_{\rm hole}\)), whereas the total spin IFE, which includes additional non-doubly-resonant terms and electron-hole asymmetries, does not necessarily correlate directly with the SHC or JDOS, as summarized in Table~\ref{tab:summary}.

\subsection{Electron–hole asymmetry in the spin IFE}
\begin{figure}[!t]
    \centering
    \includegraphics[scale=0.95]{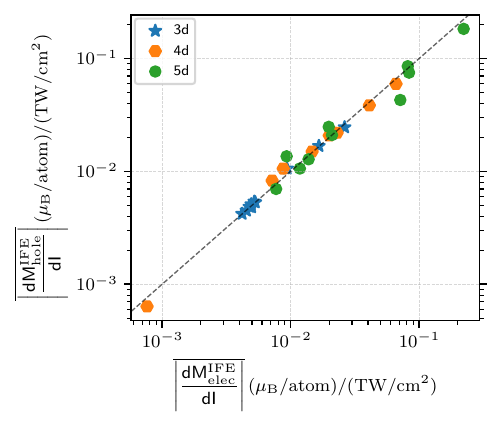}
    \caption{\label{fig:mean-ele-hole} Mean absolute values of $M^{\rm IFE}_{\rm hole}$ versus $M^{\rm IFE}_{\rm elec}$ for 3$d$, 4$d$ and 5$d$ metals in the 1$-$2~eV frequency range. Both axes are shown on a logarithmic scale. The black dashed line represents the line of equality and serves as a visual guide for comparing the electron and hole contributions.}
\end{figure}

$M^{\rm IFE}_{\rm elec}$ represents the spin moments generated by optically excited electrons under CP light, while $M^{\rm IFE}_{\rm hole}$ corresponds to spin moments from the optical de-excitation process involving holes. Together, these components define the total spin contributions to the IFE. To examine their dependence on the valence electron filling, we plot the mean absolute values of $M^{\rm IFE}_{\rm elec}$ and $M^{\rm IFE}_{\rm hole}$ in Fig.~\ref{fig:mean-ele-hole}. For most elements, these two quantities scale proportionally, indicating a strong correlation between electron and hole contributions. 

Differences between electron and hole spin moments arise from several interrelated factors, including SOI, effective mass asymmetries~\cite{Fuseya2015}, impurity scattering~\cite{Bai2015, Dale2022}, and interband transition dynamics~\cite{Karplus1954}. These effects are especially 
pronounced in materials with strong SOI. For instance, it is reported that heavy elements like bismuth, variations in band structure, and effective mass significantly affect the Landé $g$ factor~\cite{Fuseya2015}. Even in 3$d$ transition metals, where SOI is weaker, the $d$-orbital configuration plays a key role in spin-orbit-driven responses, as evidenced by the systematic variation of SHC with $d$-orbital filling~\cite{Du2014}.

In our study, 5$d$ elements such as Os and Pt exhibit the highest IFE, attributed to their strong SOI, $d$-state occupancy near the Fermi level, and strong electron-hole asymmetry. Interestingly, while 3$d$ series show similar magnitudes of $M^{\rm IFE}_{\rm elec}$ due to similar strength of SOI, the difference between $M^{\rm IFE}_{\rm elec}$ and $M^{\rm IFE}_{\rm hole}$ varies widely, highlighting the critical role of $d$-electron configurations in determining IFE response.

\begin{figure}[!t]
    \centering
    \includegraphics[scale=0.95]{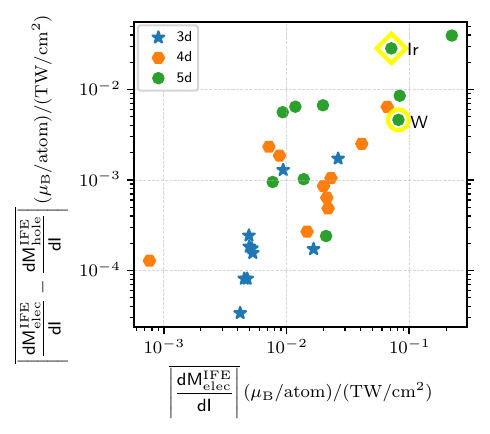}
    \caption{\label{fig:mean-ele-tot} Mean absolute values of the asymmetry between excited electron and hole spin magnetic moments, plotted against the mean absolute values of electron spin moments for 3$d$, 4$d$, and 5$d$ transition metals in the 1$-$2~eV frequency range}.
\end{figure}

\begin{table*}
\footnotesize
\begin{tabular}{c c c c c c c c c c}
\hline \hline
Element & Crystal & \multicolumn{2}{c}{\hspace*{2.0mm} Lattice parameters \hspace*{2.0mm} } & $\sigma_{\rm xy}^{\rm spin z}$ & \multicolumn{5}{c}{\hspace*{5.0mm} IFE [$\times 10^{-3} (\mu_{\rm B} / {\rm atom}) / ({\rm TW} / {\rm cm}^2 )$] \hspace*{5.0mm} } \\
symbol & structure & $a$ (\AA) & $c$ (\AA) & [$\times 10^2 \, (\hbar/e) \, S/cm$] & $M^{\rm IFE}_{\rm elec}$ & $M^{\rm IFE}_{\rm hole}$ & $M^{\rm IFE}_{\rm elec} - M^{\rm IFE}_{\rm hole}$ & $M^{\rm IFE}_{\rm ndr}$ & $M^{\rm IFE}$ \\ \\
\hline
Sc & hcp & 3.31 & 5.27 &  0.135  & \hspace{0.5mm}  4.532  \hspace{0.5mm} & \hspace{1.0mm} -4.493  \hspace{0.5mm} &  0.039 &  0.014 &  0.053 \\
Ti & hcp & 2.95 & 4.69 & -0.085  & \hspace{0.5mm} -4.185  \hspace{0.5mm} & \hspace{1.0mm}  4.213  \hspace{0.5mm} &  0.028 & -0.034 & -0.006 \\
V  & bcc & 3.03 &      & -0.441  & \hspace{0.5mm} -4.859  \hspace{0.5mm} & \hspace{1.0mm}  4.859  \hspace{0.5mm} &  0.081 & -0.054 &  0.027 \\
Cr & bcc & 2.91 &      & -1.410  & \hspace{0.5mm} -16.632 \hspace{0.5mm} & \hspace{1.0mm} 16.804  \hspace{0.5mm} &  0.172 & -0.040 &  0.132 \\
Mn & bcc & 2.79 &      & -0.705  & \hspace{0.5mm} -4.069  \hspace{0.5mm} & \hspace{1.0mm}  4.312  \hspace{0.5mm} &  0.243 & -0.052 &  0.191 \\
Fe & bcc & 2.87 &      &  2.928  & \hspace{0.5mm} -4.254  \hspace{0.5mm} & \hspace{1.0mm}  4.095  \hspace{0.5mm} & -0.159 & -0.095 & -0.254 \\
Co & hcp & 2.51 & 4.07 &  1.264  & \hspace{0.5mm}  9.393  \hspace{0.5mm} & \hspace{1.0mm} -10.553 \hspace{0.5mm} & -1.160 &  0.511 & -0.649 \\
Ni & fcc & 3.52 &      &  14.244 & \hspace{0.5mm} 26.317  \hspace{0.5mm} & \hspace{1.0mm} -24.598 \hspace{0.5mm} &  1.719 & -1.328 &  0.391 \\
Cu & fcc & 3.61 &      &  0.908  & \hspace{0.5mm}  5.305  \hspace{0.5mm} & \hspace{1.0mm} -5.318  \hspace{0.5mm} & -0.012 & -0.204 & -0.216 \\
Zn & hcp & 2.66 & 4.95 & -0.258  & \hspace{0.5mm} -5.218  \hspace{0.5mm} & \hspace{1.0mm}  5.392  \hspace{0.5mm} &  0.175 & -0.144 &  0.030 \\

Y  & hcp & 3.64 & 5.73 & 0.559  & \hspace{0.5mm}   23.20 \hspace{0.5mm} & \hspace{1.0mm} -22.048  \hspace{0.5mm} &  0.972 & -0.304 &  0.668 \\
Zr & hcp & 3.23 & 5.15 & -1.145 & \hspace{0.5mm} -21.754 \hspace{0.5mm} & \hspace{1.0mm}  21.877  \hspace{0.5mm} &  0.123 & -0.219 & -0.097 \\
Nb & bcc & 3.30 &      & -0.887 & \hspace{0.5mm} -14.747 \hspace{0.5mm} & \hspace{1.0mm}  15.015  \hspace{0.5mm} &  0.268 & -0.075 &  0.192 \\
Mo & bcc & 3.15 &      & -2.493 & \hspace{0.5mm} -21.333 \hspace{0.5mm} & \hspace{1.0mm}  21.970  \hspace{0.5mm} &  0.637 &  0.235 &  0.872 \\
Tc & hcp & 2.74 & 4.39 & -0.731 & \hspace{0.5mm} -5.747  \hspace{0.5mm} & \hspace{1.0mm}   8.035  \hspace{0.5mm} &  2.287 & -0.052 &  2.236 \\
Ru & hcp & 2.71 & 4.28 &  0.081 & \hspace{0.5mm}  8.763  \hspace{0.5mm} & \hspace{1.0mm} -10.601  \hspace{0.5mm} & -1.838 &  0.365 & -1.472 \\
Rh & fcc & 3.80 &      &  8.810 & \hspace{0.5mm} 40.997  \hspace{0.5mm} & \hspace{1.0mm} -38.492  \hspace{0.5mm} &  2.506 &  0.428 &  2.934 \\
Pd & fcc & 3.89 &      & 10.780 & \hspace{0.5mm} 60.055  \hspace{0.5mm} & \hspace{1.0mm} -59.633  \hspace{0.5mm} &  6.422 & -1.939 &  4.484 \\
Ag & fcc & 4.09 &      &  0.610 & \hspace{0.5mm}  0.764  \hspace{0.5mm} & \hspace{1.0mm} -0.636   \hspace{0.5mm} &  0.128 & -0.152 & -0.024 \\
Cd & hcp & 2.98 & 5.62 & -0.187 & \hspace{0.5mm} -19.983 \hspace{0.5mm} & \hspace{1.0mm} 20.840   \hspace{0.5mm} &  0.857 & -0.459 &  0.397 \\

La & hcp & 3.50 & 5.55 &  2.961 & \hspace{0.5mm}  83.514 \hspace{0.5mm} & \hspace{0.5mm} -74.979  \hspace{0.5mm} &  8.535 & -0.455 & 8.080 \\
Hf & hcp & 3.20 & 5.05 & -1.676 & \hspace{0.5mm} -19.780 \hspace{0.5mm} & \hspace{0.5mm}  24.815  \hspace{0.5mm} &  5.034 & -1.374 & 3.661 \\
Ta & bcc & 3.30 &      & -1.394 & \hspace{0.5mm} -13.809 \hspace{0.5mm} & \hspace{0.5mm}  12.787  \hspace{0.5mm} & -1.022 & -0.115 & -1.137 \\
W  & bcc & 3.17 &      & -7.908 & \hspace{0.5mm} -81.573 \hspace{0.5mm} & \hspace{0.5mm}  85.365  \hspace{0.5mm} &  3.792 &  3.585 & 7.377 \\
Re & hcp & 2.76 & 4.46 & -1.337 & \hspace{0.5mm}  9.893  \hspace{0.5mm} & \hspace{0.5mm} -3.434   \hspace{0.5mm} &  6.459 &  1.061 & 7.520 \\
Os & hcp & 2.73 & 4.32 & -0.366 & \hspace{0.5mm}  6.796  \hspace{0.5mm} & \hspace{0.5mm} -12.411  \hspace{0.5mm} & -5.614 &  2.678 & -2.936 \\
Ir & fcc & 3.84 &      &  3.060 & \hspace{0.5mm} 71.406  \hspace{0.5mm} & \hspace{0.5mm} -42.860  \hspace{0.5mm} & 28.547 & -1.410 & 27.137 \\
Pt & fcc & 3.92 &      & 22.859 & \hspace{0.5mm} 222.402 \hspace{0.5mm} & \hspace{0.5mm} -182.946 \hspace{0.5mm} & 39.455 &  1.686 & 41.142 \\
Au & fcc & 4.08 &      &  3.907 & \hspace{0.5mm} 7.709   \hspace{0.5mm} & \hspace{0.5mm} -6.994   \hspace{0.5mm} & 0.716  & -1.627 & -0.911 \\
\multirow{2}{*}{Hg} & \multirow{2}{*}{rhombohedral} & 4.47 & 3.04 & \multirow{2}{*}{0.802} 
& \multirow{2}{*}{20.993} & \multirow{2}{*}{-21.017} & \multirow{2}{*}{-0.024} & 
\multirow{2}{*}{-0.683} & \multirow{2}{*}{-0.707} \\
 &  & $b = 4.09$ \AA, & $\gamma = 84.4^\circ$ &  &  &  &  &  &  \\
\hline \hline
\end{tabular}
\caption{\label{tab:summary} List of 30 elements considered in this work, along with their lattice parameters, SHC, mean values of individual contributions to the IFE, and total spin IFE in the 1$-$2~eV frequency range. SHC values are scaled by a factor of $10^{-2}$, and IFE values are scaled by $10^{3}$.}
\end{table*}

We compare the asymmetry between the doubly-resonant electron and hole spin moments, defined as ($M^{\rm IFE}_{\rm elec}-M^{\rm IFE}_{\rm hole}$), with the individual contributions from electrons ($M^{\rm IFE}_{\rm elec}$). A similar analysis can be extended to the hole term ($M^{\rm IFE}_{\rm hole}$) as well. As shown in Fig.~\ref{fig:mean-ele-tot}, for most metals, this asymmetry differs significantly from $M^{\rm IFE}_{\rm elec}$ alone. Notably, in 3$d$ metals, although the magnitudes of $M^{\rm IFE}_{\rm elec}$ are relatively similar, the asymmetry varies substantially across elements.

To illustrate this effect, we highlight two representative 5$d$ metals: W and Ir. For W, the electron and hole contributions are nearly equal (0.082 and 0.085~$(\mu_{\rm B} / {\rm atom}) / ({\rm TW} / {\rm cm}^2 )$, respectively), resulting in a small total IFE. In contrast, Ir shows a larger disparity—0.071 for electrons and 0.043 for holes—leading to an IFE nearly an order of magnitude higher. This comparison underscores a key result: the greater the asymmetry between electron and hole spin moments, the larger the resulting IFE.

\subsection{\label{sec:non-doubly-res} Comparison of doubly-resonant and non-doubly-resonant contributions to IFE}

\begin{figure*}[!hbt]
    \centering
    \includegraphics[scale=0.95]{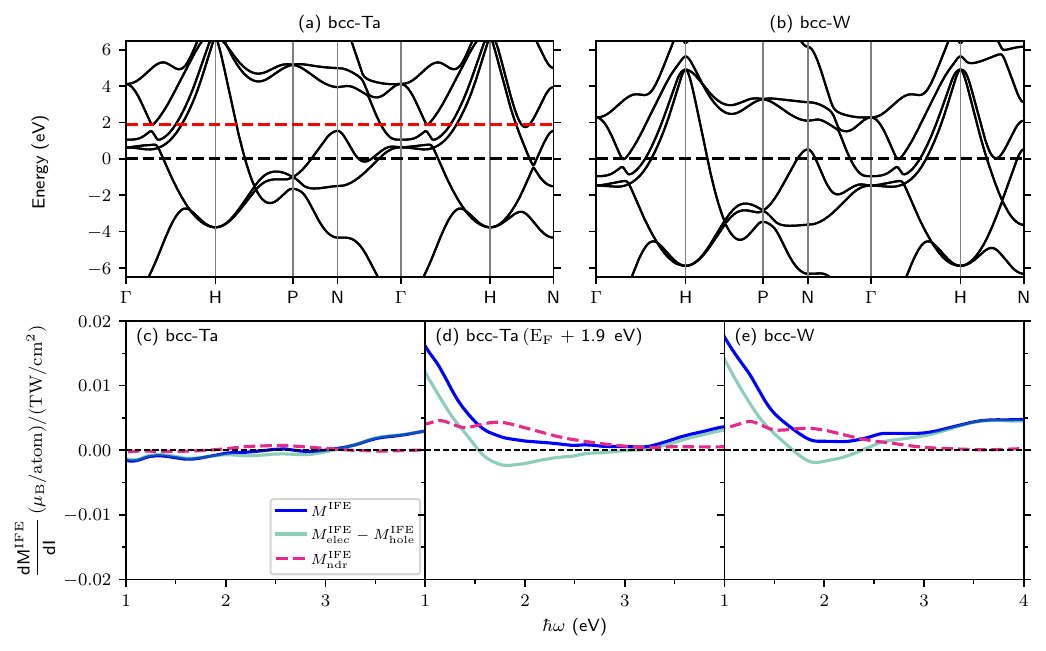}
    \vspace{-0.1cm}
    \caption{\label{fig:ta-ef-vary} (a) and (b) Band structures of bcc-Ta and bcc-W, respectively. The black dashed line indicates the Fermi level ($E_{\rm F}$), while the red dashed line in (a) marks the $E_{\rm F}$ of Ta artificially shifted by 1.9~eV. Calculated values of the total IFE ($M^{\rm IFE}$), the asymmetry in doubly-resonant terms ($M^{\rm IFE}_{\rm elec}-M^{\rm IFE}_{\rm hole}$), and the non-doubly-resonant term ($M^{\rm IFE}_{\rm ndr}$) for (c) Ta at its original $E_{\rm F}$, (d) Ta with $E_{\rm F}$ shifted by 1.9~eV, and (e) W at its original $E_{\rm F}$.}
\end{figure*}

\begin{figure}
    \centering
    \includegraphics[scale=0.95]{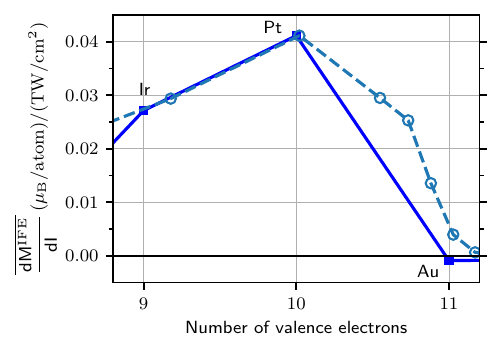}
    \vspace{-0.1cm}
    \caption{\label{fig:pt-mean-ife-total} Comparison of mean of absolute IFE for fcc-Pt as a function of electron filling, achieved by shifting the $E_{\rm F}$. Light blue open circles represent the IFE values for Pt at artificially shifted $E_{\rm F}$, while blue solid squares correspond to the IFE values of Ir, Pt, and Au at their respective actual $E_{\rm F}$.}
\end{figure}

In Eq.~\eqref{eq:ife_all_three}, the total IFE is decomposed into contributions from doubly-resonant terms, $M^{\rm IFE}_{\rm elec}$ and $M^{\rm IFE}_{\rm hole}$, and the non-doubly-resonant term, \(M^{\rm IFE}_{\rm ndr}\). To illustrate their relative importance, we analyze four representative 5\(d\) transition metals W, Os, Ir, and Pt having large IFE responses. As shown in Fig.~\ref{fig:non-and-resonant} (Appendix~\ref{app:conts-IFE}), the non-doubly-resonant contributions (detailed in Appendix~\ref{app:nonresonant}) are typically 10–100 times smaller than the doubly-resonant terms, highlighting the dominant role of the doubly-resonance in determining the total IFE.

To understand the origin of this dominance we examine the resonance conditions discussed earlier in Sec.~\ref{sec:theory}. In particular, 5$d$ transition metals exhibit strong SOI and significant electron-hole asymmetry, both of which contribute to enhancing the IFE. A related question is whether SOI, specifically through the spin matrix elements \( \braket{\phi_{n,\sigma} | \mathbf{S} | \phi_{m,\sigma'}} \) for \( n \neq m \), contributes significantly to the non-doubly-resonant term \( M_{\rm ndr} \). To test this, we isolated the effects of energy resonance and spin matrix elements. First, we set the energy denominators to a constant value (i.e., \( E_m - E_n \pm \hbar \omega + i \eta = 1 \)) to eliminate resonance effects. Under this condition, the relative magnitudes of the doubly- and non-doubly-resonant terms remained similar (see Fig.~\ref{fig:cont-IFE} in Appendix~\ref{app:conts-IFE}), confirming that the discrepancy between two contributions primarily arises from resonance conditions in the band structure. Second, to assess the role of spin matrix elements, we replaced them with a constant value ($\braket{\phi_{n{\bm k}} | M^{\rm spin} | \phi_{m{\bm k}}} = 1$). The resulting IFE was found to be one to two orders of magnitude smaller than the full calculation, indicating that spin matrix elements alone can not enhance the IFE without the presence of resonant conditions.

\subsection{\label{sec:band-struct}Band structure engineering and IFE}

To further explore the IFE trend with valence electrons, we computed the IFE while systematically shifting the Fermi levels of various transition metals, with similar crystal structures. Since the periodic arrangement of these elements is governed by $d$- and $s$-orbital filling—key factors in both crystal structure formation~\cite{Skriver1985} and IFE variation—this approach enables direct comparison.

For bcc structures, we examined Ta and W, which have similar hopping parameters and nearly identical band structures [see Figs.~\ref{fig:ta-ef-vary}(a) and~\ref{fig:ta-ef-vary}(b)]. The primary difference lies in electron densities and Fermi level ($E_{\rm F}$) position (black dashed lines). Consequently, their IFEs differ significantly [see Figs.~\ref{fig:ta-ef-vary}(c) and~\ref{fig:ta-ef-vary}(e)]. When the $E_{\rm F}$ of Ta is shifted upward by 1.9~eV to match W electron count, the resulting IFE closely resembles that of W [Fig.~\ref{fig:ta-ef-vary}(d)]. The decomposition into doubly-resonant and non-doubly-resonant terms are also similar, with minor deviations near $\hbar \omega$ = 2.5$-$4.0~eV attributed to small differences in lattice constants and SOI strength.

For fcc structures, a similar trend is observed among Ir, Pt, and Au. By rigidly shifting the Fermi level of Pt, its IFE aligns well with that of Ir (9 valence electrons) and Au (11 valence electrons), as shown in Fig.~\ref{fig:pt-mean-ife-total}. A comparable result is also found in hcp structures, where Tc and Ru exhibit similar IFE behavior upon Fermi level adjustment (Fig.\ref{fig:tc-ef-vary}, Appendix~\ref{app:IFE-hcp}). These results highlight the importance of band structure and Fermi level positioning in tuning the spin IFE—practically achievable through chemical doping or alloying.

In addition to the band structure asymmetries and SOI discussed above, electronic correlations and Hund’s coupling can subtly influence the IFE. Coulomb interactions have been shown to enhance the effective SOI, orbital moments, and spin dynamics in correlated multiorbital systems~\cite{deMedici2011, Liu2023}. For the IFE, recent Floquet-based analyses demonstrate that in strongly correlated Mott insulators, electronic correlations modify the coupling between light and spin degrees of freedom~\cite{Banerjee2022}. Hund’s coupling, by stabilizing high-spin configurations and suppressing orbital fluctuations, further enhances correlation effects and can impact spin-dependent optical transitions. Although such many-body effects are expected to be moderate in the 3$d$, 4$d$, and 5$d$ transition metals studied here, they may still contribute to quantitative variations in the IFE magnitude. A more detailed treatment incorporating beyond-DFT many-body corrections would be an interesting direction for future work.

\begin{figure*}[!t]
    \centering
    \includegraphics[scale=0.95]{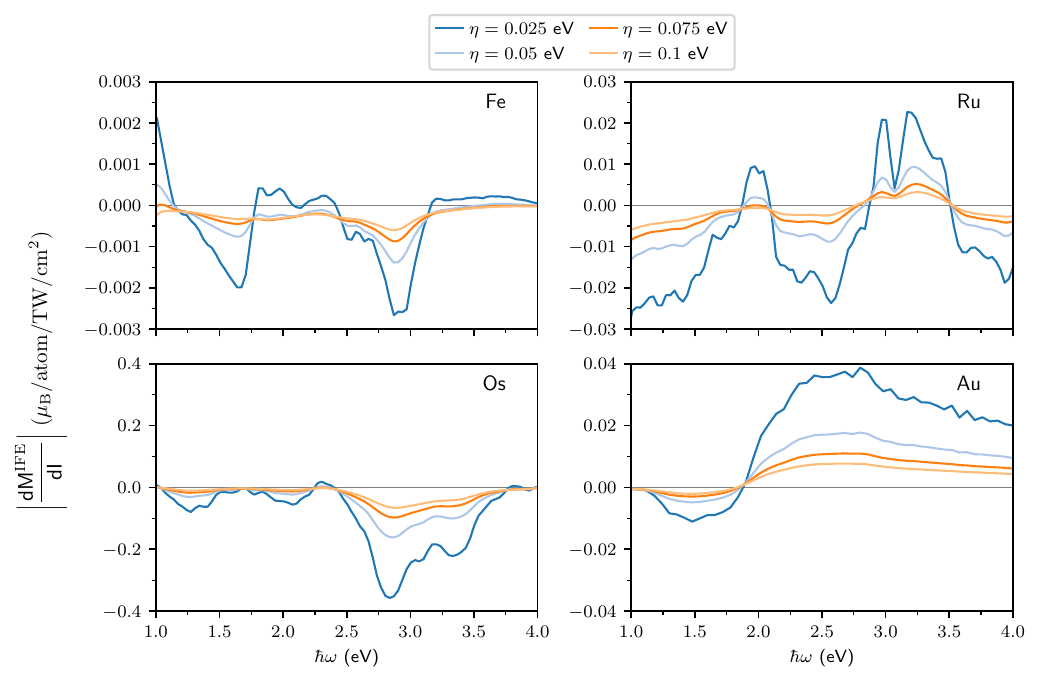}
    \vspace{-0.25cm}
    \caption{\label{fig:eta_vary}Comparison of total spin IFE in Fe, Ru, Os, and Au for varying broadening parameters. The total spin IFE is shown for selected elements (Fe, Ru, Os, and Au) with different values of the broadening parameter \(\eta\), ranging from 25 to 100 meV.}
\end{figure*}

\section{\label{app:eta-vary}Role of electron lifetime}

We now briefly comment on the importance of the electron lifetime ($\eta$) magnitude used in our calculation. Our results shown so far were obtained with $\eta=0.1$~eV. In Fig.~\ref{fig:eta_vary}, we show the calculated value of the total spin IFE for selected metals such as bcc Fe, hcp Ru, hcp Os, and fcc Au using $\eta$ set to 0.025, 0.05, 0.075, and 0.1 eV. Comparing the IFE at these lifetimes, we observe that the basic features of the IFE remain the same, but the overall magnitude is reduced near the resonance. This is consistent with the expectation that a larger $\eta$ would lead to a less pronounced resonance structure of the IFE.

\section{\label{sec:summary}Summary}

We investigated the spin contributions to the IFE in 3$d$, 4$d$, and 5$d$ transition metals. The IFE originates from asymmetries between the spin magnetic moments of photoexcited electrons and holes, driven by differences in their band structures and the influence of SOI. Greater asymmetry leads to stronger IFE responses. We observed that doubly-resonant electron contributions to the IFE ($M^{\rm IFE}_{\rm elec}$) follow a trend similar to the SHC, with both sign and magnitude depending on the number of valence electrons. However, the total spin IFE, which includes subtle asymmetries and non-doubly-resonant terms, does not directly correlate with SHC or the JDOS. A full summary of SHC, total IFE, and its components is provided in Table~\ref{tab:summary}.

Among all elements, Pt shows the highest spin IFE in the 1$-$2~eV range, while Os leads in the 2$-$4~eV range. We also demonstrated elements with similar crystal structures exhibit comparable IFE responses when their Fermi levels are aligned. These results highlight the role of valence electron filling and band structure in shaping the spin IFE and offer a foundation for future studies. 

\begin{acknowledgments}
The author acknowledges discussions with S.~Coh and R. Wilson. Computational resources were provided by the High-Performance Computing Center (HPCC) at UCR. 
\end{acknowledgments}

\appendix
\begin{figure*}[!hbt]
    \centering
    \includegraphics[scale=0.9]{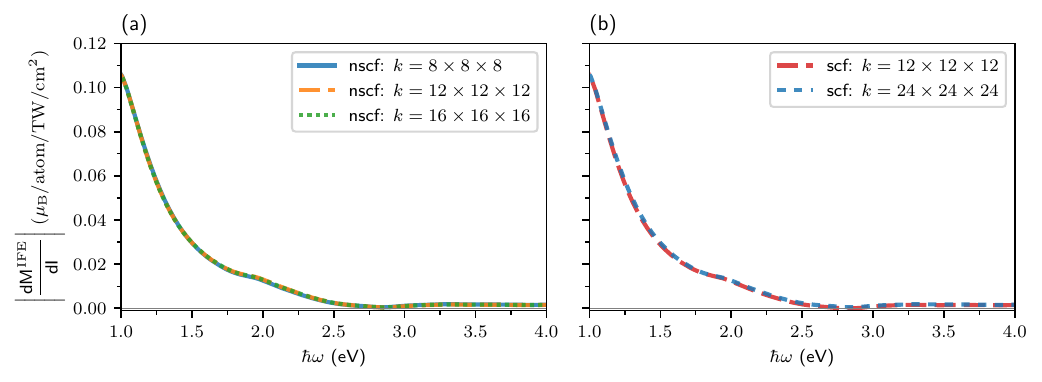}
    \vspace{-0.35cm}
    \caption{\label{fig:grid-conv} (a) Convergence of the spin IFE in Pt with varying NSCF \(k\)-grids. (b) Convergence of the IFE in Pt for two different SCF \(k\)-point grids using a fixed NSCF grid of $8\times 8 \times 8$. An electron lifetime broadening of $\eta$= 0.1~eV is used in both cases.}
\end{figure*}

\begin{widetext}
\section{\label{app:nonresonant} Non-doubly-resonant contributions to the IFE}
The expressions for interband non-doubly resonant contribution to the IFE are given by~\cite{Shashi2023},
\begin{align}
\label{eq:ndr} 
M^{\rm IFE}_{\rm ndr}
&=
M^{\rm IFE}_{\rm ndr,a}
+M^{\rm IFE}_{\rm ndr,b}
+M^{\rm IFE}_{\rm ndr,c}
+M^{\rm IFE}_{\rm ndr,d}  
-M^{\rm IFE}_{\rm ndr,e}
-M^{\rm IFE}_{\rm ndr,f}
-M^{\rm IFE}_{\rm ndr,g}
-M^{\rm IFE}_{\rm ndr,h}.
\end{align}

The eight contributions to $M^{\rm IFE}_{\rm ndr}$ are,
\begin{align}
M^{\rm IFE}_{\rm ndr,a}
&
=
\int_{\rm BZ} 
\frac{d^3k}{(2\pi)^3} 
\sum_n^{\rm occ} 
\sum_{N=1}^2
\sum_m^{\rm emp}
\sum_{M=1}^2
\sum_{m'}^{\rm emp}
\sum_{M'=1}^2
\left( 1-\delta_{m m'}\right)
\frac{
\braket{\phi_{nN} | V | \phi_{mM}} 
\braket{\phi_{mM} | M^{\rm spin} | \phi_{m'M'}}
\braket{\phi_{m'M'} | V^{\dagger} | \phi_{nN}} }
{
(E_m - E_n - \hbar \omega - i \eta)
(E_{m'} - E_n - \hbar \omega + i \eta)
} \label{eq:ndr-a}
\\
M^{\rm IFE}_{\rm ndr,b}
&
=
\int_{\rm BZ} 
\frac{d^3k}{(2\pi)^3} 
\sum_n^{\rm occ} 
\sum_{N=1}^2
\sum_m^{\rm emp}
\sum_{M=1}^2
\sum_{m'}^{\rm emp}
\sum_{M'=1}^2
\frac{
\braket{\phi_{nN} | V^{\dagger} | \phi_{mM}} 
\braket{\phi_{mM} | M^{\rm spin} | \phi_{m'M'}}
\braket{\phi_{m'M'} | V | \phi_{nN}} }
{
(E_m - E_n + \hbar \omega - i \eta)
(E_{m'} - E_n + \hbar \omega + i \eta)
}\label{eq:ndr-b}
\\
M^{\rm IFE}_{\rm ndr,c}
&
=
\int_{\rm BZ} 
\frac{d^3k}{(2\pi)^3} 
\sum_n^{\rm occ} 
\sum_{N=1}^2
\sum_m^{\rm emp}
\sum_{M=1}^2
\sum_{m'}^{\rm emp}
\sum_{M'=1}^2
2 {\rm Re}
\frac{
\braket{\phi_{nN} | M^{\rm spin} | \phi_{mM} }
\braket{\phi_{mM} | V | \phi_{m'M'}}
\braket{\phi_{m'M'} | V^{\dagger} | \phi_{nN}} }
{
(E_m - E_n + 2 i \eta)
(E_{m'} - E_n - \hbar \omega + i \eta)
}\label{eq:ndr-c}
\\
M^{\rm IFE}_{\rm ndr,d}
&
=
\int_{\rm BZ} 
\frac{d^3k}{(2\pi)^3} 
\sum_n^{\rm occ} 
\sum_{N=1}^2
\sum_m^{\rm emp}
\sum_{M=1}^2
\sum_{m'}^{\rm emp}
\sum_{M'=1}^2
2 {\rm Re}
\frac{
\braket{\phi_{nN} | M^{\rm spin} | \phi_{mM} }
\braket{\phi_{mM} | V^{\dagger} | \phi_{m'M'}}
\braket{\phi_{m'M'} | V | \phi_{nN}} }
{
(E_m - E_n + 2 i \eta)
(E_{m'} - E_n + \hbar \omega + i \eta)
} \label{eq:ndr-d}
\\
M^{\rm IFE}_{\rm ndr,e}
&
=
\int_{\rm BZ} 
\frac{d^3k}{(2\pi)^3} 
\sum_n^{\rm occ} 
\sum_{N=1}^2
\sum_{n'}^{\rm occ}
\sum_{N'=1}^2
\sum_m^{\rm emp}
\sum_{M=1}^2
\left( 1-\delta_{n n'}\right)
\frac{
\braket{\phi_{nN} | V | \phi_{mM}} 
\braket{\phi_{mM} | V^{\dagger} | \phi_{n'N'}}
\braket{\phi_{n'N'} | M^{\rm spin} | \phi_{nN}} }
{
(E_m - E_n - \hbar \omega - i \eta)
(E_m - E_{n'} - \hbar \omega + i \eta)
} \label{eq:ndr-e}
\\
M^{\rm IFE}_{\rm ndr,f}
&
=
\int_{\rm BZ} 
\frac{d^3k}{(2\pi)^3} 
\sum_n^{\rm occ} 
\sum_{N=1}^2
\sum_{n'}^{\rm occ}
\sum_{N'=1}^2
\sum_m^{\rm emp}
\sum_{M=1}^2
\frac{
\braket{\phi_{nN} | V^{\dagger} | \phi_{mM}} 
\braket{\phi_{mM} | V | \phi_{n'N'}}
\braket{\phi_{n'N'} | M^{\rm spin} | \phi_{nN}} }
{
(E_m - E_n + \hbar \omega - i \eta)
(E_m - E_{n'} + \hbar \omega + i \eta)
}\label{eq:ndr-f}
\\
M^{\rm IFE}_{\rm ndr,g}
&
=
\int_{\rm BZ} 
\frac{d^3k}{(2\pi)^3} 
\sum_n^{\rm occ} 
\sum_{N=1}^2
\sum_{n'}^{\rm occ}
\sum_{N'=1}^2
\sum_m^{\rm emp}
\sum_{M=1}^2
2 {\rm Re}
\frac{
\braket{\phi_{nN} | M^{\rm spin} | \phi_{mM} }
\braket{\phi_{mM} | V^{\dagger} | \phi_{n'N'}}
\braket{\phi_{n'N'} | V | \phi_{nN}} }
{
(E_m - E_n + 2 i \eta)
(E_m - E_{n'} - \hbar \omega + i \eta)
} \label{eq:ndr-g}
\\
M^{\rm IFE}_{\rm ndr,h}
&
=
\int_{\rm BZ} 
\frac{d^3k}{(2\pi)^3} 
\sum_n^{\rm occ} 
\sum_{N=1}^2
\sum_{n'}^{\rm occ}
\sum_{N'=1}^2
\sum_m^{\rm emp}
\sum_{M=1}^2
2 {\rm Re}
\frac{
\braket{\phi_{nN} | M^{\rm spin} | \phi_{mM} }
\braket{\phi_{mM} | V | \phi_{n'N'}}
\braket{\phi_{n'N'} | V^{\dagger} | \phi_{nN}} }
{
(E_m - E_n + 2 i \eta)
(E_m - E_{n'} + \hbar \omega + i \eta)
}. \label{eq:ndr-h}
\end{align}
\end{widetext}

\section{\label{app:grid-conv} IFE convergence with respect to NSCF and SCF \(k\)-point grids}

\begin{figure*}[!hbt]
    \centering
    \includegraphics[width=7.0in]{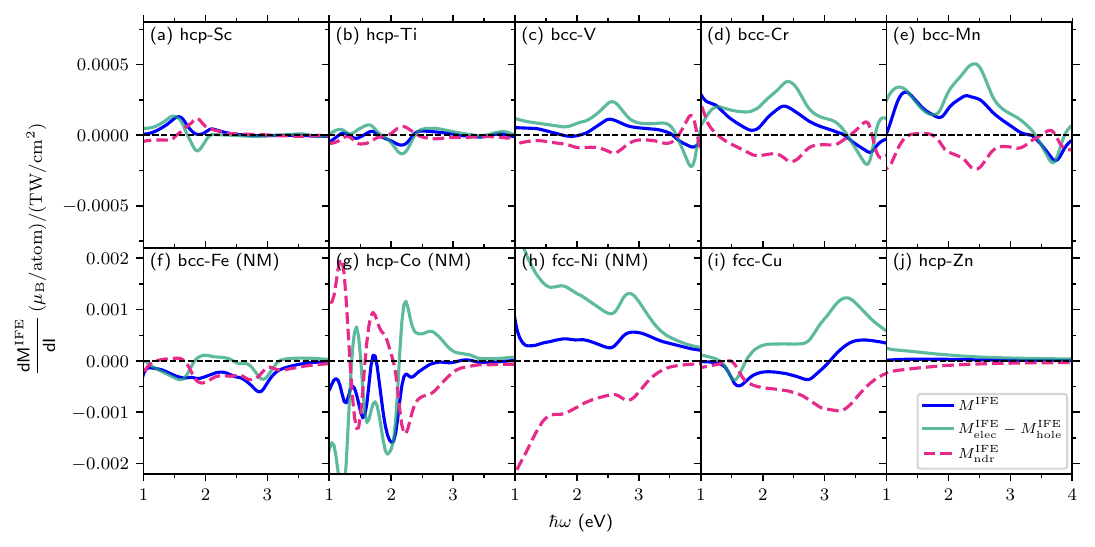}
    \vspace{-0.15cm}
        \caption{\label{fig:3d-metals} Frequency dependence of the spin IFE ($M^{\rm IFE}$) for 3$d$ metals. The decomposition into doubly-resonant electron-hole asymmetry ($M^{\rm IFE}_{\rm elec} - M^{\rm IFE}_{\rm elec}$) and non-doubly resonant ($M^{\rm IFE}_{\rm ndr}$) contributions are also shown. In this study, we assumed nonmagnetic (NM) structures for Fe, Co, and Ni, resulting in double degenerate bands.} 
\end{figure*} 

\begin{figure*}[!t]
    \centering
    \includegraphics[width=7.0in]{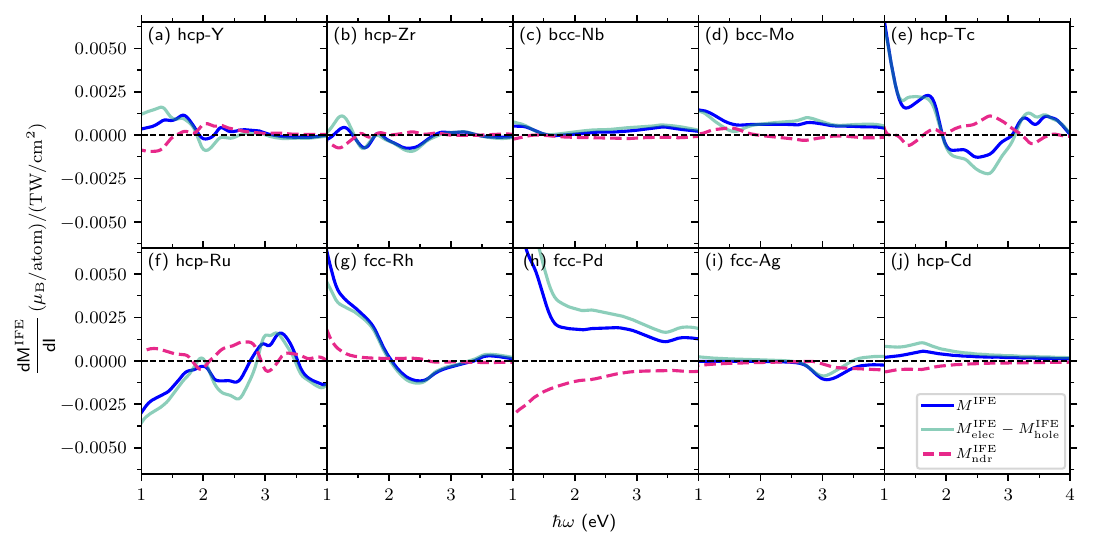}
    \vspace{-0.15cm}
    \caption{\label{fig:4d-metals} Frequency dependence of $M^{\rm IFE}$ and asymmetry term $M^{\rm IFE}_{\rm elec}- M^{\rm IFE}_{\rm hole}$, and $M^{\rm IFE}_{\rm ndr}$ for 4$d$-metals.}
\end{figure*}

\begin{figure*}[!t]
    \centering
    \includegraphics[width=7.0in]{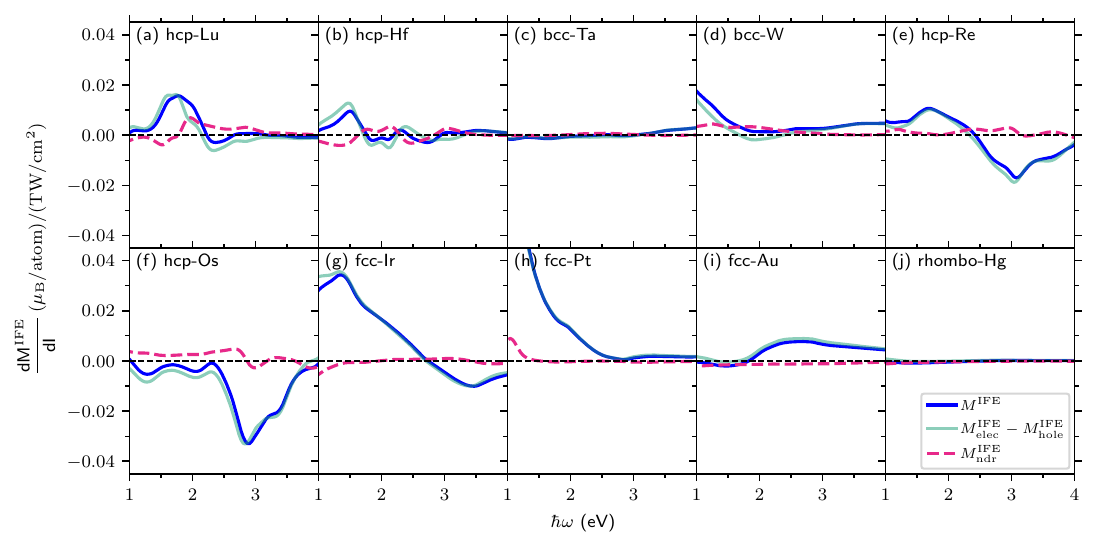}
    \vspace{-0.35cm}
    \caption{\label{fig:5d-metals} Frequency dependence of $M^{\rm IFE}$ and asymmetry term $M^{\rm IFE}_{\rm elec}- M^{\rm IFE}_{\rm hole}$, and $M^{\rm IFE}_{\rm ndr}$ for 5$d$-metals.} 
\end{figure*}

\begin{figure*}[!hbt]
    \centering
    \includegraphics[scale=0.9]{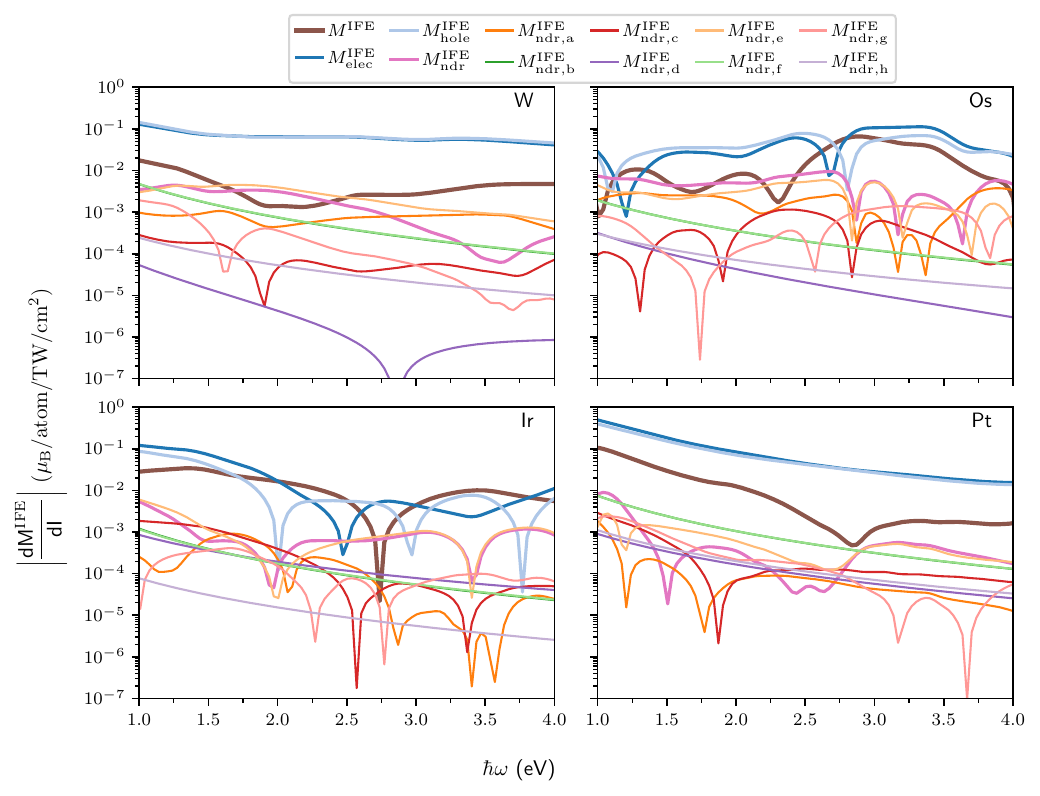}
    \vspace{-0.25cm}
    \caption{\label{fig:non-and-resonant} Comparison of interband contributions to the IFE in selected 5\(d\) transition metals. The doubly-resonant terms (\(M^{\rm IFE}_{\rm elec}\) and \(M^{\rm IFE}_{\rm hole}\)) and all terms in the non-doubly-resonant contributions (\(M^{\rm IFE}_{\rm ndr}\)) for bcc W, hcp Os, fcc Ir, and fcc Pt are shown. The vertical axis is plotted on a logarithmic scale. A broadening parameter of \(\eta = 0.1\) eV is used.}
\end{figure*}

Figure~\ref{fig:grid-conv}(a) illustrates the convergence of the IFE response in Pt as the \(\mathbf{k}\)-point meshes used in the NSCF calculations increase from \(8 \times 8 \times 8\) to \(12 \times 12 \times 12\) and \(16 \times 16 \times 16\). The results show that the spin IFE magnitude remains unchanged with this denser \(\mathbf{k}\)-meshes, indicating convergence with respect to the NSCF grids.

To further investigate, we analyzed the IFE by reducing the SCF \(\mathbf{k}\)-mesh from \(24 \times 24 \times 24\) to \(12 \times 12 \times 12\), while keeping the NSCF grid fixed at \(8 \times 8 \times 8\). As shown in Fig.~\ref{fig:grid-conv}(b), these coarser SCF meshes still yield sufficiently accurate charge densities. However, we found the IFE curve to be slightly different for Au with SCF \(\mathbf{k}\)-mesh of \(12 \times 12 \times 12\) (not shown here). Therefore, for consistency, we employed the denser SCF \(\mathbf{k}\)-meshes of \(24 \times 24 \times 24\) across all fcc and bcc structures, throughout our study.

\section{\label{app:IFE-with-freq}IFE as a function of frequency of light}

\begin{figure*}[!hbt]
    \centering
    \includegraphics[scale=0.9]{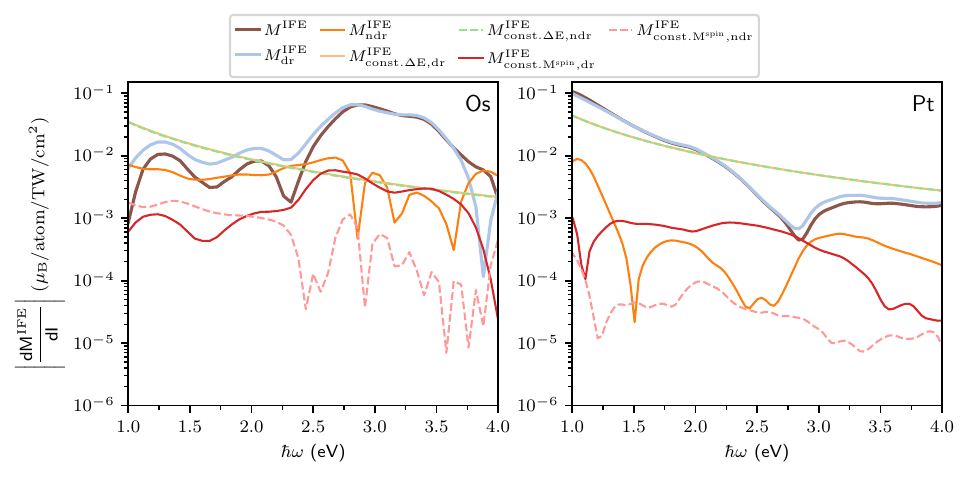}
    \vspace{-0.25cm}
    \caption{\label{fig:cont-IFE} Effect of energy resonance and spin matrix elements on the IFE response in Pt and Os. Shown are the total IFE (\(M^{\rm IFE}\)), the doubly-resonant electron-hole contribution (\(M^{\rm IFE}_{\rm elec} - M^{\rm IFE}_{\rm hole} \equiv M_{\rm dr}^{\rm IFE}\)), and the non-doubly-resonant term (\(M_{\rm ndr}^{\rm IFE}\)). The terms \(M_{\rm const.\,\Delta E,\,dr}^{\rm IFE}\) and \(M_{\rm const.\,\Delta E,\,ndr}^{\rm IFE}\) represent the IFE contributions with the energy denominators set to unity in the doubly- and non-doubly-resonant terms, respectively. Similarly, \(M_{\rm const.\,M^{\rm spin},\,dr}^{\rm IFE}\) and \(M_{\rm const.\,M^{\rm spin},\,ndr}^{\rm IFE}\) are obtained by replacing the spin matrix elements with a constant value (\(\braket{\phi_{n\bm{k}} | M^{\rm spin} | \phi_{m\bm{k}}} = 1\)) in the respective terms.}
\end{figure*}
\begin{figure*}[!hbt]
    \centering
    \includegraphics[scale=0.97]{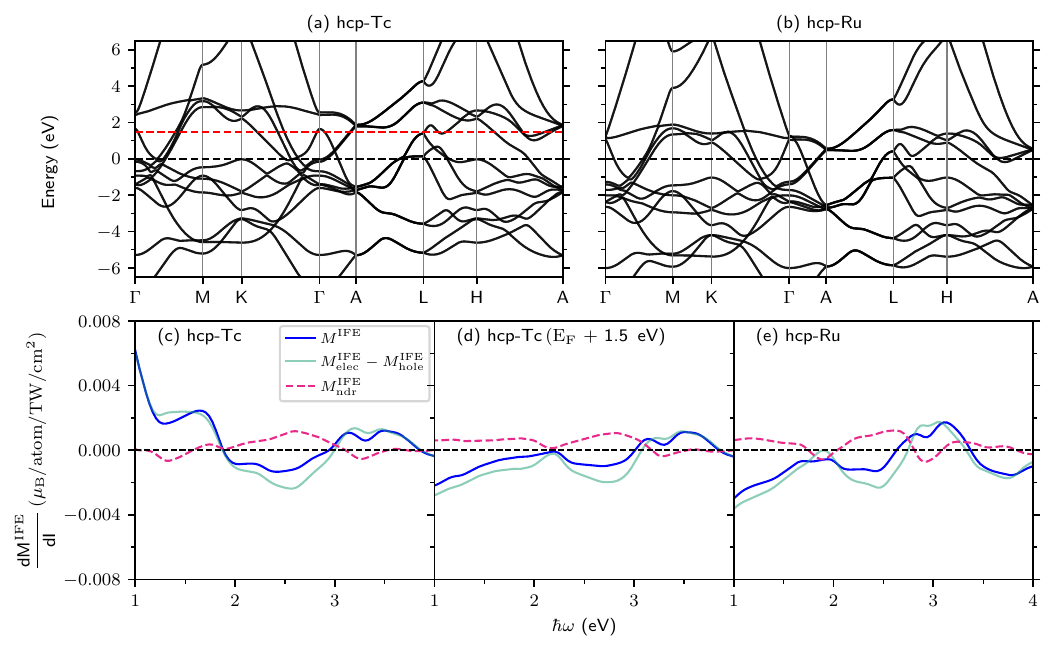}
    \vspace{-0.25cm}
    \caption{\label{fig:tc-ef-vary} Band structures and IFE response in hcp-Tc and hcp-Ru. (a) and (b) show the band structures of hcp-Tc and hcp-Ru, respectively, with their actual Fermi levels (\(E_{\rm F}\)) represented by black dashed lines. In panel (a), the red dashed line indicates the Fermi level of Tc shifted upward by 1.5 eV. Panels (c)–(e) display the calculated IFE and its components for: (c) Tc at its actual \(E_{\rm F}\), (d) Tc with \(E_{\rm F}\) shifted by 1.5 eV, and (e) Ru at its actual \(E_{\rm F}\).}
\end{figure*}

Figures~\ref{fig:3d-metals}$-$\ref{fig:5d-metals} show spin IFE and its decomposition into doubly-resonant ($M^{\rm IFE}_{\rm elec}-M^{\rm IFE}_{\rm hole}$) and non-doubly-resonant ($M^{\rm IFE}_{\rm ndr}$) components for 3$d$, 4$d$ and 5$d$ metals, arranged by atomic number. 

In 3$d$ series, for ferromagnetic Fe, Co, and Ni, hypothetical non-magnetic structures were used. Contributions from doubly-resonant and non-doubly-resonant terms nearly cancel for Zn resulting in negligible IFE. In some cases, such as fcc Ni and Cu, the non-doubly-resonant term is comparable in magnitude to the doubly-resonant term. Cu, in particular, shows an enhanced response in the 3$-$4 eV range due to $d \rightarrow s$ interband transitions. Early 3$d$ elements like Sc and Ti exhibit negligible $M^{\rm IFE}$ due to low $d$ electron density. Cr, with half-filled $d$ orbitals, shows large values of both $M^{\rm IFE}_{\rm elec}$ and $M^{\rm IFE}_{\rm hole}$, resulting in low asymmetry and total IFE. In 4$d$ metals, IFE increases with valence electrons up to Pd (with nearly filled $d$ and $s$ orbitals), then decreases for Ag and Cd. In 5$d$ elements, the non-doubly-resonant term is typically smaller, resulting in higher overall IFE values due to stronger SOI. Pt shows the highest spin IFE, followed by Ir in the 1$-$2~eV range, while Os exhibits the largest IFE response in the 2$-$4~eV range. Fully-filled systems, such as Hg, show negligible IFE, consistent with filled-shell behavior in Zn (3$d$) and Cd (4$d$). 

\section{\label{app:conts-IFE} Analysis of doubly-resonant and non-doubly-resonant terms to IFE} 

Figure~\ref{fig:cont-IFE} compares the IFE response when energy resonance effects are suppressed (constant energy approximation, \(E_m - E_n \pm \hbar \omega + i\eta = 1\)) and when spin matrix elements are replaced with a constant value (\(\braket{\phi_{n{\bm k}} | M^{\rm spin} | \phi_{m{\bm k}}} = 1\)). The results show that setting constant energy preserves the relative dominance of doubly-resonant and non-doubly-resonant terms, confirming that energy resonance is the primary driver of the IFE. In contrast, replacing spin matrix elements with a constant value significantly reduces the overall IFE response, demonstrating that spin matrix elements alone contribute minimally. These findings highlight that while spin-orbit interaction (SOI) influences matrix elements, the dominant mechanism governing the IFE response in 5\(d\) elements is the energy resonance condition.

\section{\label{app:IFE-hcp} IFE in hcp metals by Fermi level tuning}
For hcp crystals, we demonstrate the equivalence of IFE between Tc and Ru, by rigidly shifting the Fermi levels of Ru by 1.5~eV as shown in Figure~\ref{fig:tc-ef-vary}. It shows that the IFE of Tc is similar to that of Ru with the shifted Fermi level.

%

\end{document}